\documentstyle[psfig]{mn2e}
\input epsf

\def\kms{\,km~s$^{-1}$}      

\def\deg{\hbox{$^\circ$}}

\def\lesssim{\mathrel{\hbox{\rlap{\hbox{\lower4pt\hbox{$\sim$}}}\hbox{$<$}}}}
\def\gtrsim{\mathrel{\hbox{\rlap{\hbox{\lower4pt\hbox{$\sim$}}}\hbox{$>$}}}}

\newcommand{\mdot}{\mbox{$\stackrel{.}{\textstyle M}$}}

\begin{document}

\title[The nature of QU Car]
{QU Car: a very high luminosity nova-like binary with a carbon-enriched 
companion}

\author[Janet E. Drew et al. ]
{ Janet E. Drew$^1$, Louise E. Hartley$^1$, Knox S. Long$^2$ \& Johan van der 
Walt$^3$\\
$^1$Imperial College of Science, Technology and Medicine,
Blackett Laboratory, Prince Consort Road, London,  SW7 2BW, U.K.\\
$^2$Space Telescope Science Institute, 3700 San Martin Drive, Baltimore,
MD 21218, USA.\\
$^3$ Space Research Unit, Department of Physics, Potchefstroom University, 
Potchefstroom 2520, South Africa.}

\date{received,  accepted}

\maketitle
\begin{abstract}
QU~Car is listed in cataclysmic variable star catalogues as a nova-like
variable.  This little-studied, yet bright interacting binary is
re-appraised here in the light of new high-quality ultraviolet (UV) 
interstellar line data obtained with STIS on board the Hubble Space 
Telescope.  The detection of a component of interstellar absorption at a 
mean LSR velocity of $-$14 km s$^{-1}$ indicates that the distance to QU~Car 
may be $\sim$2~kpc or more -- a considerable increase on the previous 
lower-limiting distance of 500~pc.  If so, the bolometric luminosity of 
QU~Car could exceed $10^{37}$ ergs~s$^{-1}$.  This would place this binary 
in the luminosity domain occupied by known compact-binary supersoft 
X-ray sources.  Even at a 500~pc, QU~Car appears to be the most luminous 
nova-like variable known.  

New intermediate dispersion optical spectroscopy of QU~Car spanning 
3800--7000~\AA\ is presented.   These data yield the discovery that 
C{\sc iv}~$\lambda\lambda$5801,12 is present as an unusually prominent 
emission line in an otherwise low-contrast line spectrum.  Using measurements 
of this and other lines in a recombination line analysis, it is shown that 
the C/He abundance as proxied by the n(C$^{4+}$)/n(He$^{2+}$) ratio may be as 
high as 0.06 (an order of magnitude higher than the solar ratio).  
Furthermore, the C/O abundance ratio is estimated to be greater than 1.  
These findings suggest that the companion in QU Car is a carbon star.  If so, 
it would be the first example of a carbon star in such a binary.
An early-type R star best matches the required abundance pattern and
could escape detection at optical wavelengths provided the distance to
QU~Car is $\sim$2~kpc or more.
\end{abstract}


\section{Introduction}

  QU~Car was first described in the astronomical literature by Schild 
(1969), after it had been noted as a potentially-interesting emission line 
object by Stephenson, Sanduleak \& Schild (1968).  According to Schild, this 
variable resembled an old nova spectroscopically and exhibited irregular 
short-term brightness fluctuations not exceeding 0.2 amplitudes in amplitude.
Over a decade later Gilliland \& Phillips (1982, herafter GP82) 
presented high time resolution blue spectra of QU~Car that they were able to 
analyse for radial velocity variation.  They determined a period of 
10.9 hours from centroid motion in He{\sc ii}~$\lambda$4686 and H$\beta$ line 
emission, thereby confirming the binary nature of the system.  They also 
argued from the absence of detectable secondary star absorption lines and the 
implied apparent magnitude limit of the secondary star that QU Car was at a 
distance of at least 500~pc (see also Duerbeck 1999).  The designation of 
QU~Car as a nova-like variable in catalogues of cataclysmic variables (CV) 
originates with GP82.  Warner (1995, Table 4.1) further qualifies this by 
placing it in the UX~UMa sub-class.  The physical model for UX~UMa systems
is that these are white-dwarf interacting binaries in which the mass transfer 
rate is high enough ($3\times10^{-9} < \mdot < 1\times10^{-8}$ 
M$_{\odot}$~yr$^{-1}$) to sustain the accretion disk in an opaque 
high state. 

  The blue spectrum of QU~Car is quite remarkable in two respects.
First, the contrast of all line features against the continuum is very low.
The highest contrast emission lines peak at under 1.2 times the continuum
level.  Secondly, and uniquely among known CV, emission in the 
carbon/nitrogen blend centred near 4650~\AA\ is comparable in equivalent 
width to that of the He{\sc ii}~$\lambda$4686 emission.  Impressed by this 
unusual property, GP82 deconvolved the 4650~\AA\ emission and showed that it 
is dominated by carbon emission.  This was an important finding because it 
consigned fluorescent excitation via the Bowen mechanism to a minor role, 
and pointed instead toward significant CN abundance enhancement as the 
explanation for the feature's great relative strength.

\begin{figure*}
\begin{picture}(0,270)
\put(0,0){\includegraphics{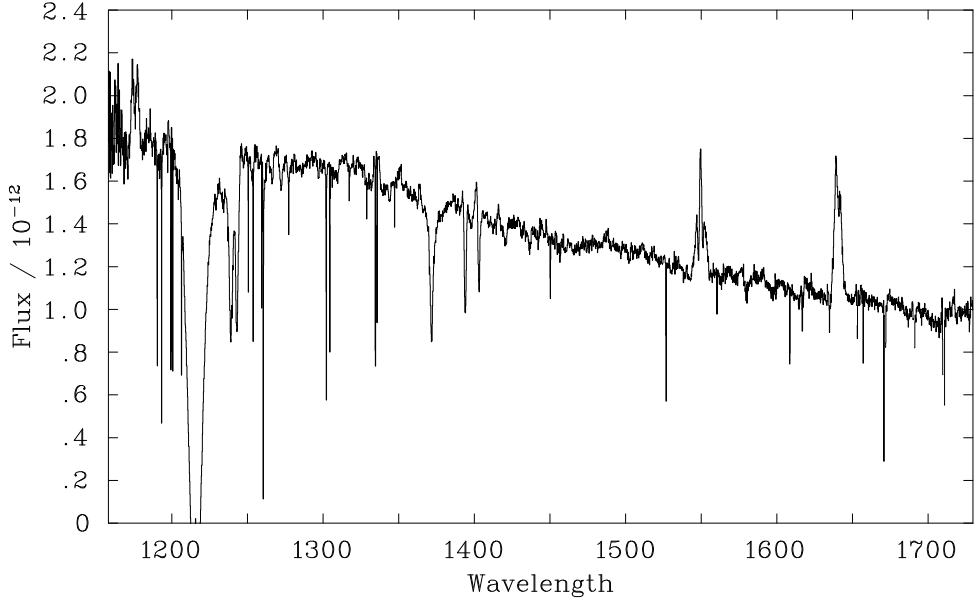}}
\end{picture}
\caption{The HST/STIS ultraviolet spectrum of QU Car.  The data shown
are the grand sum of all the echelle data obtained to date (Hartley et
al. 2002).}
\label{fig:hst}
\end{figure*}

  QU Car was observed a number of times at UV wavelengths with
the International Ultraviolet Explorer (IUE).  In particular, a series
of observations was obtained with a view to searching for orbital-phase
linked variability in the stronger UV resonance lines (Knigge et al 1994),
a phenomenon often apparent in high-state CV.  Nothing conclusive was 
found, but it was noted that O{\sc v}~$\lambda$1371 absorption was unusually
prominent in the spectrum.  As a relatively bright source at a visual 
magnitude of between 11.1 and 11.5 (see Table 4.1 in Warner 1995), it was 
natural to include QU~Car in a recent Hubble Space Telescope (HST) programme 
of high time- and spectral-resolution UV 
spectroscopy aimed at testing a model for the accretion disk winds seen in 
nova-like variables (and in QU~Car -- see Knigge et al 1994).  With the 
advantage of both the high spectral resolution and high signal delivered by 
the E140M echelle of the Space Telescope Imaging Spectrograph (STIS) on
this target, it nevertheless became all too plain that the UV spectrum of 
QU~Car deviates from the nova-like variable norm.  The main peculiarities
are the higher than usual ionization signalled by bright 
He{\sc ii}~$\lambda$1640 emission and the (already noted) strong 
O{\sc v}~$\lambda$1371 absorption, together with the weak mass loss 
signatures in e.g. N{\sc v}~$\lambda$1240 and C{\sc iv}~$\lambda$1549.  
An in-depth discussion of the HST/STIS E140M data is to be found in Hartley, 
Drew \& Long (2002).

  The aim of this work on QU~Car is to re-open the thus-far sparse 
discussion on the nature of this binary.  Our conclusion will be that
its bland classification as a nova-like variable has served to conceal an
extraordinary binary: new 
evidence, derived both from the HST/STIS observations and from the first 
broad-band optical spectrum of QU~Car to be obtained in the digital era, 
suggests instead that this object is unusually luminous (with a mass
transfer rate perhaps as high as $\gtrsim 10^{-7}$ M$_{\odot}$ yr$^{-1}$) 
and that the accreting material is also significantly carbon-enriched.

In Section 2, we present the interstellar line
spectrum as it emerges from the co-add of all the HST/STIS E140M 
data available (shown as Figure 1).  This is used to argue that the distance 
to QU~Car could be $\sim$2~kpc or more.  In Section 3 we estimate the
bolometric luminosity both for the near distance of 500~pc and the
representative longer one of 2~kpc.  Then in Section 4 we compare
and contrast the optical spectrum of QU~Car with those of a selection of 
supersoft sources presented by Cowley et al (1998).  Finally, we begin 
the task of analysing the optical emission line spectrum for element 
abundances and find, intriguingly, that carbon is significantly 
enhanced (Section 5).  We close with a discussion in which we draw attention 
to the puzzle the apparent abundance peculiarity poses for the stellar 
content of this short-period binary.

\section{The reddening, the UV interstellar lines and the distance to 
QU~Car}

\begin{figure*}
\begin{picture}(0,240)
\put(0,0){\includegraphics{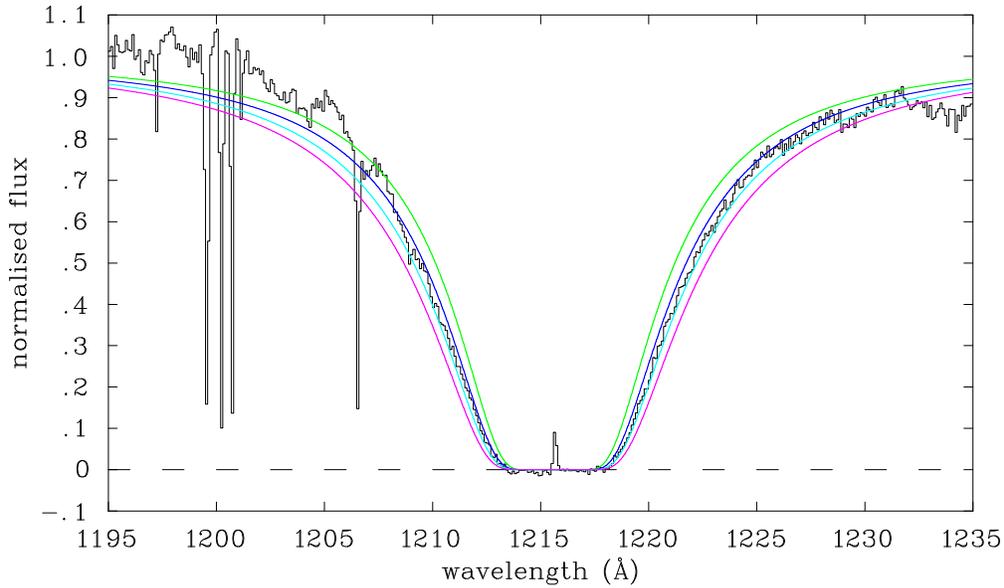}}
\end{picture}
\caption{A comparison between theoretical Ly$\alpha$ damping profiles and 
the normalised merged spectrum.  Profiles are drawn for column densities 
$N(H{\sc i})/(10^{20} {{\rm cm}}^{-2}) = 5$, $6$, $7$ and $8$. A variety of 
continuum normalisations were tried.  Here we show the result of the simplest 
normalisation strategy: a local linear interpolation after fixing the 
continuum level at 1171\AA\ and 1262\AA .  It is evident that an 
$N(H{\sc i})$ column of between $6\times 10^{20}$ and $7\times 10^{20}$ 
cm$^{-2}$ gives the best match to the data.  If a quadratic or power law 
continuum fit is used instead, the only difference is stronger favouring of 
$N(H{\sc i}) = 6\times 10^{20}$ cm$^{-2}$. The excess flux relative to the
fit at $\sim 1200$~\AA\ is because we have placed the continuum under the
assumption that there is unidentified line emission at these wavelengths
(cf. fig. 1).}
\label{fig:hst_la}
\end{figure*}

As bright CV are usually nearby objects, the interstellar reddening towards 
them is normally negligible, and interstellar absorption lines in their 
spectra are weak.  QU~Car, however, is known 
to exhibit a measureable reddening even though, at $m_v \sim 11.4$, only
3 nova-like variables are listed by Warner (1995) as brighter.  On the basis 
of correcting the UV spectral energy distribution (SED) for the 2200~\AA\ 
broad interstellar extinction feature, Verbunt (1987) determined 
$E_{B-V} = 0.1 \pm 0.03$ for this object.  This makes QU Car one of only 11 
out of Verbunt's sample of 51 CV (observed with IUE) to be reddened by this 
amount or more.  We have re-examined IUE archive data and can confirm this 
reddening estimate, 
noting that the quoted error bound is conservative ($3\sigma$).  

    The HST/STIS (1160-1700 \AA) data now available to us provide the means 
to obtain an independent, if indirect, check on the reddening.  We can 
measure the neutral hydrogen column density by comparing the shape of the 
observed Ly$\alpha$ damping wings with theoretical predictions.  
For this object, with its relatively high degree of ionization, it is 
reasonable to assume that the Ly$\alpha$ damping wings are interstellar in 
origin.  The extraction of the data we used for this purpose was one that 
included corrections for echelle inter-order scattered light, so that the
flux minimum in the Ly$\alpha$ profiles is zero as it should be.  The results 
of our comparison are shown as Figure 2.  We conclude from this that the 
neutral hydrogen column density is $(6 \pm 1) \times 10^{20}$ cm$^{-2}$.  For 
a standard gas-to-dust ratio of $N(H{\sc i})/E_{B-V} = 5.8\times 10^{21}$ 
mag$^{-1}$ cm$^{-2}$ (Bohlin, Savage \& Drake 1978), this column converts to 
$E_{B-V} = 0.1 \pm 0.015$ -- in agreement with the pre-existing UV SED-based 
reddening determination.  It seems $E_{B-V} = 0.1$ 
towards QU~Car can be regarded as a good working estimate.

   The higher than typical reddening for a CV is accompanied by a richer than 
usual interstellar line spectrum.  Indeed we have been able to identify and
obtain measurements of 34 transitions in the sum of the HST/STIS echelle
observations (Figure 1, see also Hartley et al 2002).  The results of these
measurements are presented as Table 1.  For every line, an equivalent
width is reported.  In most cases we are also able to provide meaningful 
velocity measurements.  The clearly saturated absorption lines are listed in 
the top part of the table. The velocities given (to the nearest \kms ) for 
all but two of these are intended to convey an impression of the range
over which absorption occurs.  Specifically they are the `full width at 
half-flux' points -- the velocities where the observed flux is half that in 
the adjacent continuum.  It is very noticeable that the absorption in these 
strong lines is displaced towards negative LSR velocities, suggesting the 
blending within them of more than one velocity component.

\begin{figure}
\begin{picture}(0,270)
\put(0,0){\includegraphics{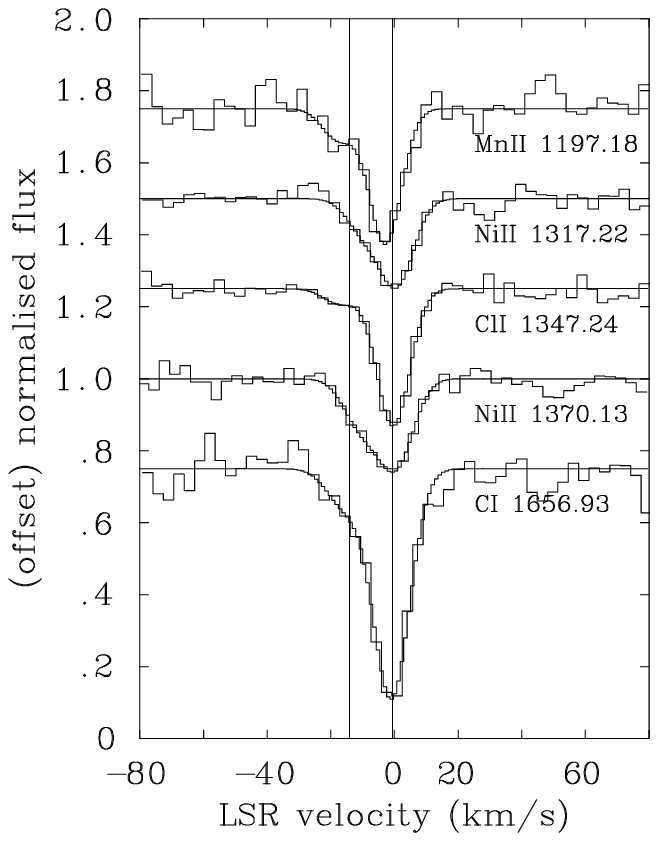}}
\end{picture}
\caption{Ultraviolet interstellar lines from the co-added HST/STIS spectrum
of QU Car.  The lines shown here are unsaturated at the instrumental 
resolution of the E140m echelle grating ($R \sim 23000$).  In each there is
clear evidence of a second blueshifted velocity component.  The smooth 
lines superposed on the binned profile data are the Gaussian fits giving the
velocities listed for these transitions in Table 1.}
\label{fig:is}
\end{figure}

\begin{table*}
\caption{Equivalent width and velocity measurements of the interstellar lines 
in the ultraviolet spectrum of QU~Car. Where the equivalent is quoted to the
nearest 5~m\AA , it was determined by approximate integration over the entire
line profile; where it is quoted to the nearest 0.1 m\AA , it has been
determined by fitting 1 or more Gaussian components to the line profile. The
velocity data given on heavily saturated line profiles, in the top part of the
table, are the half-continuum points (where the profile passes through 50 
percent of the continuum level). Otherwise the given velocities and errors 
are derived from Gaussian fits to the line profiles.  In the `appearance' 
column, the label `s' indicates a heavily saturated line, `$\sim$s' indicates 
close to saturation, while the labels 1 and 2 respectively indicate 
unsaturated profiles presenting with one and two velocity components.
}
\label{t_cont}
\begin{tabular}{lrcrcrc}
\hline
Atomic Transition  & Equivalent    & appearance & \multicolumn{3}{c}{velocity data} & Note \\  
                   & Width (m\AA ) &            & \multicolumn{3}{c}{(in LSR km s$^{-1}$)} \\
\hline
Saturated lines: & & & & & \\
Si~{\sc ii} 1190.42 & 140 & s & $-$24 & $\rightarrow$ & $+$9 & \\
Si~{\sc ii} 1193.29 & 165 & s & $-$24 & $\rightarrow$ & $+$12 & \\
N~{\sc i} 1199.55  & 145 & s & $-$27 & $\rightarrow$ & $+$10 & \\ 
N~{\sc i} 1200.22  & 130 & s & $-$26 & $\rightarrow$ & $+$10 & \\
N~{\sc i} 1200.71  & 125 & s & $-$24 & $\rightarrow$ & $+$8 & \\
Si~{\sc iii} 1206.50 & 105 & s & $-$18 & $\rightarrow$ & $+$15 & \\
S~{\sc ii} 1250.58 & 87.3 & $\sim$s & $-(12.1\pm0.9)$ & & $+(0.8\pm0.6)$ & \\
S~{\sc ii} 1253.81 & 105 & s & $-$17 & $\rightarrow$ & $+$9 & \\
S~{\sc ii} 1259.52 & 115 & s & $-$18 & $\rightarrow$ & $+$9 & \\
O~{\sc i} 1302.17  & 190 & s & $-$25 & $\rightarrow$ & $+$17 & \\
Si~{\sc ii} 1304.37 & 135  & s & $-$20 & $\rightarrow$ & $+$9 & \\
C~{\sc ii} 1334.53 & 210 & s & $-$27 & $\rightarrow$ & $+$18 & \\
C~{\sc ii} 1335.71 & 130 & s & $-$21 & $\rightarrow$ & $+$9 & \\
Si~{\sc ii} 1526.71 & 160 & s & $-$22 & $\rightarrow$ & $+$10 & \\
Fe~{\sc ii} 1608.45 & 121.4 & $\sim$s & $-(12.2\pm0.7)$ & & $+(0.2\pm0.5)$ & \\
Al~{\sc ii} 1670.79 & 160 & s & $-$20 & $\rightarrow$ & $+$9 & \\
\multicolumn{3}{l}{Unsaturated lines:} & & & \\
S~{\sc iii} 1190.20 & 55 & ? &                 & &                & a \\
C~{\sc i} 1193.03 & 13.4 & 1 &  &  & $-(3.8\pm0.8)$ & \\
Mn~{\sc ii} 1197.18 & 23.3 & 2 & $-(17.6\pm2.1)$ & & $-(2.9\pm0.6)$ & Fig. 3 \\
Mn~{\sc ii} 1199.39 & 20 & ?   &               & &                & a \\
Mn~{\sc ii} 1201.12 & 20 & 2?   &               & &                & b \\
Mg~{\sc ii} 1239.93 & 25 & 2?  &               & &                & b \\
Mg~{\sc ii} 1240.40 & 14.7 & 1  &               & & $-(1.8\pm0.6)$ & \\
C~{\sc i} 1260.74 & 21.4 & 2? & $-(19.3\pm4.3)$ & & $+(0.7\pm0.9)$ & \\
C~{\sc i} 1277.25 & 30.8 & 1    &               & & $+(1.7\pm0.3)$ & \\
C~{\sc i} 1280.14 & 8.8  & 1    &               & & $+(2.0\pm0.6)$ & \\
P~{\sc ii} 1301.87 & 25 & 1?   &               & &                & a \\
Ni~{\sc ii} 1317.22 & 18.9 & 2 & $-(9.6\pm2.2)$ & & $+(0.7\pm0.9)$ & Fig. 3 \\
C~{\sc i} 1328.83 & 24.9 & 2? & $-(21.6\pm3.4)$ & & $-(2.3\pm0.4)$ & \\
Cl~{\sc i} 1347.24 & 24.7 & 2 & $-(17.1\pm2.7)$ & & $-(0.4\pm0.3)$ & Fig. 3 \\
Ni~{\sc ii} 1370.13 & 23.1 & 2 & $-(10.5\pm1.7)$ & & $+(0.4\pm1.0)$ & Fig. 3 \\
P~{\sc ii} 1532.53 & 13.5 & 2?   & $-(20.0\pm7.5)$ & & $-(2.5\pm1.2)$ & \\
C~{\sc i} 1560.31 & 35.1 & 1    &               & & $-(0.5\pm0.4)$ & \\
C~{\sc i} 1656.93 & 58.3 & 2  & $-(15.6\pm3.3)$ & & $-(1.0\pm0.6)$ & Fig. 3 \\
\hline
\end{tabular}

\noindent
$^a$ Feature is located in the wing of a saturated interstellar line and is 
therefore difficult to measure \\
$^b$ Both profiles may contain two components but are hard to measure
because of distorting noise or artifacts \\
\end{table*}

   Fortunately there are many weaker unsaturated interstellar lines present
in the UV spectrum.  These are listed in the lower part of Table 1.
In these cases it is of value to fit one or more Gaussian components to the 
observed profile, thereby obtaining measures of mean component velocities and 
line widths.  The latter are of no physical interest since we can expect
them to be dominated by instrumental broadening -- indeed these widths are
scattered within the range 12 to 14 km/s (implying an effective resolving
power for the combination of STIS echelle grating and aperture used of about 
23000).  Several of the weaker interstellar lines appear to comprise only one 
velocity component: these are always centred at an LSR velocity within 2 or 
3 \kms\ of zero.  However many echo the saturated lines in that there is 
evidence in them of a second, distinctly blueshifted, velocity component.  
The best examples of these are shown in Figure 3.  The weighted mean velocity 
of this second blueshifted component is $-14.5\pm1.1$ \kms .  If this mean is 
calculated including the blueshifted component velocities of the 
nearly-saturated Fe{\sc ii}~$\lambda$1608 and S{\sc ii}~$\lambda$1250 lines, 
this figure decreases a little to $-12.5\pm0.5$ \kms .
  
   If QU Car is at $\sim$500~pc, we might expect only 
to detect interstellar absorption lines centred on the Local Standard of 
Rest.  However since QU Car is located at a relatively low galactic latitude
($b = 7.7\deg $) and there is evidence of a second component of interstellar 
absorption at $\sim -14$~\kms , it is possible that the distance is 
significantly greater than 500 pc.  Indeed we can use the $\sim -14$~\kms 
component to estimate a new lower limit on the distance applying the 
assumption that this component is due to a cloud whose velocity is determined 
by the general Galactic rotation law.

   An earlier example of the use of this distance estimation method is to be 
found in Crawford \& Barlow (1991), where the distance to a Wolf-Rayet star 
(just a few degrees away from QU~Car) is shown to be large by assigning 
kinematic distances to components observed in the interstellar 
Na{\sc i} D line absorption in its spectrum.  If the same approach is used and 
we also adopt the Galactic rotation model favoured by Fich, Blitz \& Stark 
(1989), we find that the projected Galactic rotation velocity shifts in the 
negative direction moving along the line of sight to QU~Car, reaching a 
minimum of $-18.3$~\kms\ at a distance of 3.3~kpc.  A velocity of $-14$~\kms\ 
is first reached at the lesser distance of 1.8~kpc.  
This suggests a distance for QU Car of about 2 kpc.  At a distance of 
exactly 2~kpc and at $b = 7.7\deg $, QU~Car would be located 250 pc south of 
the Galactic mid-plane.  This is not problematic.  At the extreme limit, it
is clear that QU~Car is in the Galaxy: using data due to Dickey \& Lockman 
(1990), we estimate that the total H{\sc i} column through the Galaxy along
QU Car's sightline is $2.2\times10^{21}$ cm$^{-2}$, or about three times
the column out to QU~Car.

  In the absence of supporting observations of interstellar lines in the
spectra of stars of known distance in the same part of the sky as QU~Car, it
remains possible that the interstellar cloud responsible for the blueshifted
component in QU Car's spectrum is not a perfect tracer of Galactic rotation.
Accordingly, in the rest of this paper, the status of QU~Car will be evaluated
within the context of two possible distances: 500~pc and 2~kpc.

\section{The spectral energy distribution and estimates of the luminosity}

  To date, the only part of the spectral energy distribution (SED) to be
defined well enough for quantitative scrutiny is the ultraviolet range.
To begin to define the SED, we have gathered all the IUE observations
obtained of QU Car in June 1991 (5 long-wavelength and 12 short wavelength
spectra, see Knigge et al. 1994).  These have been merged and then corrected 
for reddening assuming $E_{B-V} = 0.1$.  The resultant ultraviolet spectral 
energy distribution obtained in this way is shown in Figure 4.  It is compared
with simple power laws in wavelength of the form, $F_{\lambda} \propto 
\lambda^{\gamma}$, where $\gamma$ is a constant.  Each power law shown
was normalised at a wavelength of 3000 \AA .  We find that $\gamma = -2.25$
runs through the dereddened energy distribution very convincingly and that 
the index only has to change by about 0.1 to produce a significantly worse
match.  This is not a very reddening-sensitive result either, given that the 
reddening is itself modest: if $E_{B-V}$ is lowered from 0.1 to 0.08, the 
best fitting power law index to the dereddened SED drops by just 0.1.  
The value of the power law index is consistent with that expected of an
optically-thick accretion disk.

\begin{figure*}
\begin{picture}(0,280)
\put(0,0){\includegraphics{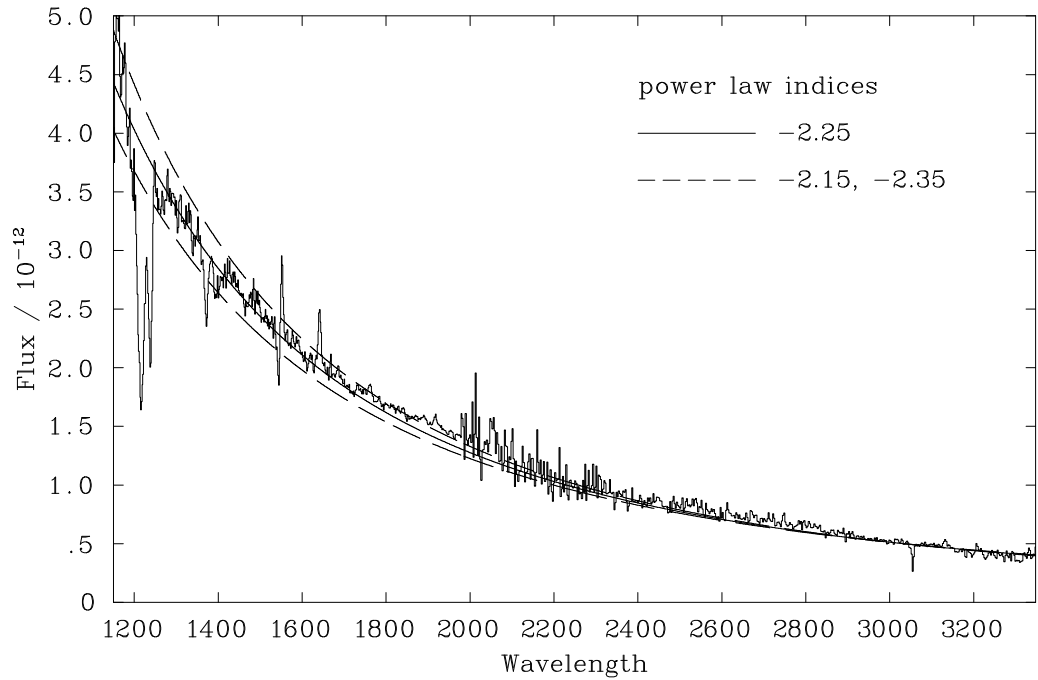}}
\end{picture}
\caption{The ultraviolet spectral energy distribution after dereddening
assuming $E_{B-V} = 0.1$.  The data are merged IUE low dispersion spectra
obtained in June 1991.  The solid curve superposed on the spectrum is
a power law with spectral index, $\gamma - -2.25$, fit to the data by eye.  
Also shown are power laws with indices 0.1 different from the best fitting 
value.}
\label{fig:is}
\end{figure*}

  At this point, we can ask whether the power law that works so well in 
the UV extrapolates satisfactorily into the optical.  It does and is
sufficient for present purposes.  The June 1991 merged IUE spectrum gives a 
dereddened flux at 3000~\AA\ of $5.1\times10^{-13}$ ergs 
s$^{-1}$ cm$^{-2}$ \AA $^{-1}$.  This extrapolates
to $1.3\times10^{-13}$ ergs s$^{-1}$ cm$^{-2}$ \AA $^{-1}$ in the $V$ band
at 5556 \AA .  Reddening this by 0.31 magnitudes and converting to a magnitude
using Hayes' (1985) zero point, we `predict' $V = 11.4$.  Fortuitously, this
is the same as the mean magnitude quoted for QU~Car on the SIMBAD database.  
In truth, and as one would expect for an object classified as a nova-like 
variable, there is photometric evidence of optical variability to within a 
few tenths of a magnitude (e.g. Schild 1968, GP82, Kern \& Bookmyer 1986, 
Hartley 2002).  Absolute photometric data suggest that QU Car has 
reached, at its brightest, $m_v = 10.9$ and, at its faintest, $m_v \simeq 12$ 
tending to spend more time in the bright half of this range (Hiltner \& 
Gordon 1971, Warner 1995, collected amateur observer data in Hartley 2002).

\begin{table}
\caption{Estimates of bolometric luminosities and equivalent mass accretion
rates for QU~Car.  Figures are given for the two distances discussed in 
Section 2.  The SED adopted in the integration to obtain the bolometric
luminosities was $F_\lambda = F_{3000}(\lambda/3000)^{-2.25}$ with $F_{3000}
= 5.13\times10^{-13}$ ergs cm$^{-2}$ s$^{-1}$ \AA $^{-1}$ (see Figure 4) with 
a short wavelength cut off at either 912\AA\ or 228\AA . The equivalent mass 
accretion rates, $\mdot_{a}$ were derived assuming accretion onto a 
1~M$_{\odot}$ white dwarf with a radius 0.01~R$_{\odot}$.}
\label{tab_lbol}
\begin{tabular}{lcc}
\hline
  & \multicolumn{2}{c}{Distance (pc)} \\
  & 500 & 2000 \\
\hline
Absolute magnitude, $M_V$ & 2.6 & -0.4 \\
 & & \\
L$_{bol}/10^{37}$ (ergs s$^{-1}$): & & \\
Power law cut off at 912\AA\ & 0.02 & 0.3 \\
Power law cut off at 228\AA\ & 0.09 & 1.5 \\
 & & \\
Equivalent $\mdot_{a}$ (M$_{\odot}$ yr$^{-1}$): & & \\
Power law cut off at 912\AA\ & $1\times10^{-8}$ & $2\times10^{-7}$ \\
Power law cut off at 228\AA\ & $8\times10^{-8}$ & $1\times10^{-6}$ \\
\hline
\end{tabular}

\end{table}

  Using the $\gamma = -2.25$ power law representation of the SED together 
with the two distance estimates, we have obtained estimates of the 
bolometric luminosity of QU Car.  These are set out in Table 
2.  At this time there are no useful observations of this binary at FUV or EUV 
wavelengths.  To deal with this lack of constraint our estimates are
calculated on the basis of what are likely to be lower-limiting and
upper-limiting assumptions regarding the FUV/EUV SED.  To represent an 
extremely conservative lower limit, Table 2 provides bolometric luminosities 
calculated on the assumption that the power law has a short wavelength 
cut-off at the Lyman limit (912 \AA ).  The alternative assumption we have 
made -- that the power law continues on down to 228~\AA , the He$^{+}$ Lyman 
edge -- can be viewed as a rough upper limit for the moment. In time we
may discover this `limit' yields estimates not too different from the correct 
figures for the following reasons.  We note that, while it is entirely 
credible that the SED has peaked 
at a wavelength longer than 228~\AA , it also must persist into the He$^{+}$ 
Lyman continuum: the presence of both N{\sc v}~$\lambda$1240 and 
O{\sc v}~$\lambda$1371 in absorption in the UV spectrum (Figure 1) certainly 
demands the supply of photons in the He$^{+}$ Lyman continuum capable of 
photo-ionising N$^{3+}$ and O$^{3+}$ from either their ground or excited 
levels (Drew 1989).  At the same time, the fact that QU Car was not 
detected by the soft X-ray sensitive R\"ontgen Satellit (ROSAT, Verbunt 
et al 1997) gives some confidence that the 228~\AA\ (54 eV) cutoff will not 
lead to significant underestimation.

  In order to give the luminosity esimates in Table 2 a more obvious physical 
meaning, they have been convered to mass accretion rates assuming a 
1~M$_{\odot}$ white dwarf accretor of radius 0.01~R$_{\odot}$.  These figures 
show how luminous QU Car is, even if, quite remarkably, no photons are 
emitted shortward of 912~\AA\ and it is at the near distance of 500~pc.  The 
equivalent mass accretion rate already exceeds $10^{-8}$ M$_{\odot}$ 
yr$^{-1}$, placing it a little higher than any accepted nova-like variable 
accretion rate (e.g. Warner 1995, p245).  Another way of expressing this is 
to note that $M_V = 2.6$ for QU~Car placed at 500~pc is over a magnitude 
higher than the nova-like variable norm of $M_V \sim 4$ (Warner 1995, 
Table 4.16).  If the distance to QU~Car is closer to 2~kpc than it is to 
500~pc, the data in Table 2 point to a luminosity within a factor of a few
of 10$^{37}$ ergs~s$^{-1}$ and an equivalent mass accretion rate in the 
vicinity of $10^{-7}$ M$_{\odot}$~yr$^{-1}$.  These orders of magnitude 
match those discussed in the context of mass transfer occurring on the thermal 
timescale in close-binary supersoft sources (Kahabka \& van den Heuvel 1997).

\section{The optical spectrum of QU Car}

   The potentially high luminosity and indeed the high ionization seen at UV
wavelengths in QU Car (see Figure 8 in Hartley et al 2002) suggests 
that an optical spectroscopic comparison with those of known close-binary 
supersoft sources might be enlightening.

The optical spectral characteristics of supersoft sources have been 
described by Cowley et al. (1998).  The spectra of the six sources they 
studied are characterized by strong He~{\sc ii}~$\lambda$4686 emission 
relative to H$\beta$, together with O~{\sc vi} line emission at 3811~\AA\ and 
at 5290~\AA .  The equivalent widths of the He{\sc ii} and H{\sc i}
line emission are up to an order of magnitude higher in these supersofts,
compared to QU~Car.  Nevertheless the (He{\sc ii}~$\lambda$4686)/H$\beta$ 
equivalent width ratio for QU~Car is not discordant with the supersoft norm:  
specifically, for the 4 objects listed in Table 5 of Cowley et al (1998), for 
which the ratio can be determined, it ranges from 1.9 up to 3.5 -- as judged
from the data of GP82, this ratio is $\gtrsim$2 in QU~Car.  A number of 
supersofts also emit in the C~{\sc iv} $\lambda\lambda$5801,12 doublet.
The usual pattern at these longer wavelengths is rough equality between the 
O~{\sc vi} $\lambda$5290 and He~{\sc ii} $\lambda$5411 emission, with the 
C~{\sc iv} $\lambda\lambda$5801,12 somewhat weaker than the
He~{\sc ii} lines (see Figure 1 in Cowley et al 1998).  

Hitherto, the only published optical spectra of QU Car have been those 
obtained by GP82, spanning the wavelength range 4300-4900 \AA.  This does
not cover the range needed to provide a comparison with the O{\sc vi} and 
C{\sc iv} lines.  To rectify this problem, we have obtained intermediate 
dispersion spectra covering the much wider range, 3800 to 6800~\AA .  A log 
of the observations is given in Table 3.  The data were obtained at the 
SAAO 1.9-metre telescope, using the grating spectrograph and SITe CCD.  To 
achieve the broad wavelength coverage, data were taken at
two grating angles.  The dispersion of the data obtained using grating 6 was 
close to 1~\AA\ pixel$^{-1}$, yielding a resolution of $\sim$2 \AA .  The 
only calibration observations obtained were arc spectra for determining the 
wavelength scale -- our goal was limited to finding and identifying weak 
emission lines.  The spectra were flat-fielded, 
wavelength-calibrated and sky-subtracted using routines within the NOAO/IRAF  
package.  Coincidentally the two sets of blue spectra were obtained almost
exactly 2 orbital periods apart -- and indeed the details of the line profile
shapes do not change significantly between the two observations.  This has 
meant that it has been acceptable to directly co-add the two sets 
(5 exposures) without correcting for radial velocity changes.  The total 
count per pixel in the co-added blue spectrum ranges from 40,000 at 
3800~\AA\ to a maximum of 100,000 at 4800~\AA .  The one red exposure 
obtained yields a count that varies between 10,000 pixel$^{-1}$ at 6800~\AA\ 
and 17,000 pixel$^{-1}$ at 5600~\AA .

\begin{figure*}
\begin{picture}(0,280)
\put(0,0){\includegraphics{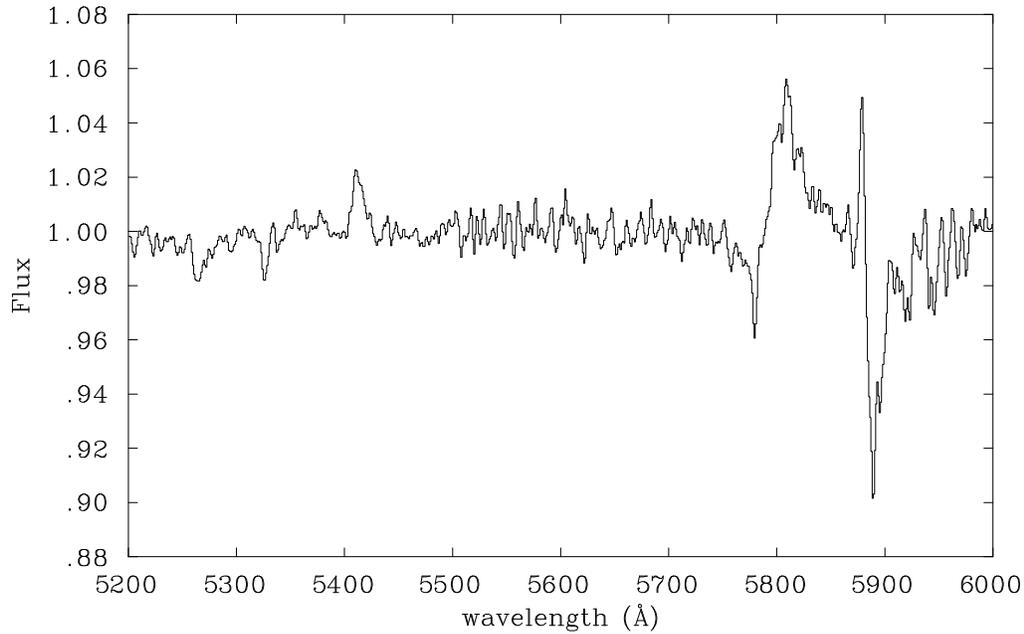}}
\end{picture}
\caption{The optical spectrum between 5200~\AA\ and 6000~\AA .  The 
separately obtained blue and red spectra overlap between 5200~\AA\ and 
5500~\AA , giving a higher signal/noise ratio in this range.  The 
prominent absorption at the red end of the range is interstellar 
Na~{\sc i} -- this partially obscures He~{\sc i}~$\lambda$5876 emission
sitting shortward of it.  The C~{\sc iv}~$\lambda\lambda$5801,12
emission is much stronger than He~{\sc ii}~$\lambda$5411.
Note the DIB at 5780~\AA\ and the hint of weak broad blueshifted absorption 
component to the C~{\sc iv} line.
}
\label{fig:civ}
\end{figure*}

\begin{table}
\caption{Journal of optical spectroscopy.
}
\label{tab_opt}
\begin{tabular}{llll}
\hline
Wavelength & Date & Start time  & Exposure \\
Range (\AA ) &    & (UT)       & (sec)    \\
\hline
3800 - 5500 & 20/02/02 & 01:45 & 600 \\
            &          & 02:00 & 900 \\ 
            &          & 23:20 & 600 \\
            &          & 23:31 & 600 \\
            &          & 23:43 & 600 \\
5000 - 6800 & 21/02/02 & 23:29 & 600 \\
\hline    
\end{tabular}
\end{table}

The spectra show no evidence of O~{\sc vi} emission.  For 
O~{\sc vi}~$\lambda$5290, our upper limit on the equivalent width is of 
order 10 m\AA . For O{\sc vi}~$\lambda$3811, our limit is 
worse -- about 50 m\AA -- due to the declining count rate and the corrugation 
of the spectrum by very weak, converging Balmer series lines.  By contrast, 
there is no difficulty at all picking out the 
C~{\sc iv}~$\lambda\lambda$5801,12 blend  -- it could even be said to tower
over the He~{\sc ii} $\lambda$5411 line in the same spectral region.
An excerpt of the spectrum containing these two lines is shown as Figure
5.  This is quite unlike the Cowley et al (1998) supersoft sources.  The blue 
wing of the C~{\sc iv} line appears to be eroded both by a narrow absorption 
that is very likely to be the 5780~\AA\ diffuse interstellar band, and 
possibly by a weak blueshifted P~Cygni absorption component.  Considering 
the lower level of the C~{\sc iv} transition lies 39~eV above the C$^{3+}$ 
ground state, this is quite remarkable, if true.

   Since it is such a high lying transition and unlikely to be flourescently
excited, it is possible that the observed emission in C~{\sc iv} 
$\lambda\lambda$5801,12 forms mainly in the C$^{4+}$ $\rightarrow$ C$^{3+}$ 
recombination cascade.  In view of the high degree of ionization already 
implied by the strong O~{\sc v}~$\lambda$1371 absorption in the UV spectrum, 
it is likely that C$^{4+}$ is the most abundant carbon ion
(the ionization potentials for O$^{3+}$ and C$^{3+}$ are respectively
77~eV and 65~eV).  Significant further ionization to C$^{5+}$ can be ruled 
out, given that the threshold for C$^{4+}$ ionization lies at 490~eV.  This 
interpretation is consistent with the observation that the profile of the 
C~{\sc iv}~$\lambda$1549 UV resonance transition is typically dominated by 
emission, rather than the P Cygni absorption that might signal C$^{3+}$ as
more abundant than C$^{4+}$ (Figure 1, Hartley et al 2002).  

   In contrast to this, the optical spectrum contains no trace of the 
highly-excited optical O~{\sc vi} lines and indeed O~{\sc v}~$\lambda$5590 
line, which would also form mainly by recombination (see e.g. Kingsburgh, 
Barlow \& Storey 1995).  This may be an indication that oxygen is not ionized 
up to either O$^{5+}$ or O$^{6+}$ (the ionization potentials of O$^{4+}$ and 
O$^{5+}$ are respectively 114~eV and 138~eV).  What of the N$^{4+}$ $-$ 
N$^{5+}$ ion balance?  If optical N~{\sc v} recombination emission can be 
identified, we will then have learned that QU~Car's SED extends to 
high enough energies to present a photon flux in the vicinity of the N$^{4+}$ 
ionization threshold at 98~eV.  The analogous transition to 
C~{\sc iv} $\lambda\lambda$5801,12 is N~{\sc v}~$\lambda\lambda$4604,20 -- 
this can be looked for, as can the even more highly-excited 
N~{\sc v}~$\lambda\lambda$4945,50 blend.  Both blends are in fact weakly 
present.  This is shown in the plot of the 4500 -- 5000 \AA\ part of the 
optical spectrum (Figure 6).  In the data of GP82, the 
N~{\sc v}~$\lambda\lambda$4604,20 blend was less evident.  
This difference is very likely related to the modest optical/UV
variability exhibited by QU~Car.  Crudely, it would seem that the high
energy limit to QU~Car's SED falls at about 100~eV.

\begin{figure*}
\begin{picture}(0,280)
\put(0,0){\includegraphics{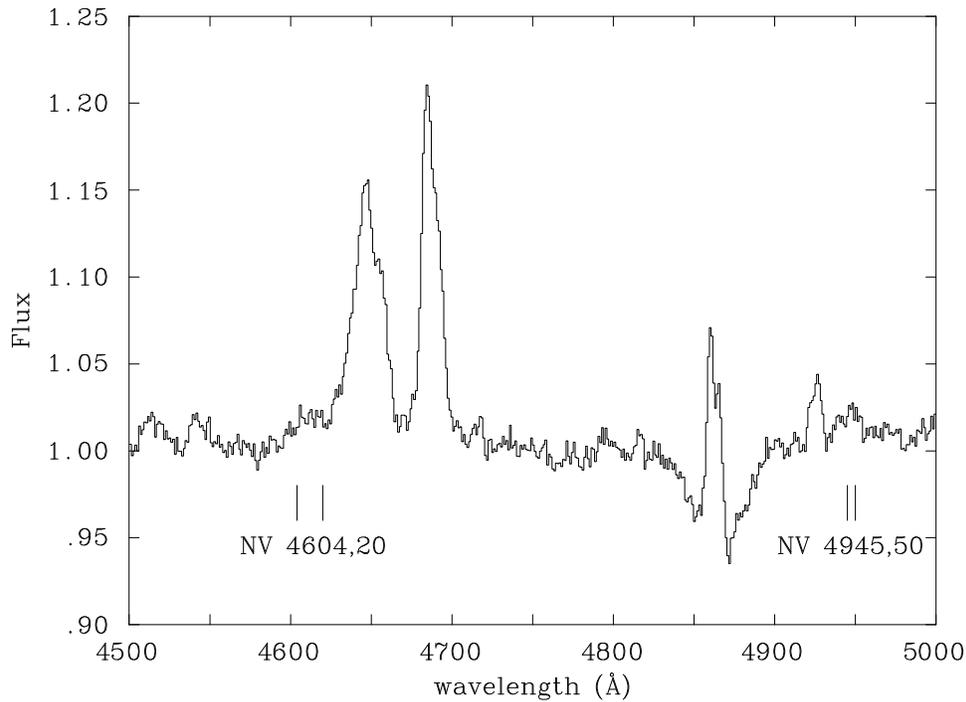}}
\end{picture}
\caption{The raw counts optical spectrum between 4500~\AA\ and 5000~\AA . 
Compared to GP82 data, the He~{\sc ii}~$\lambda$4686 emission is now taller 
than the 4650~\AA\ blend (which incorporates some hitherto unacknowledged 
C~{\sc iv}~$\lambda$4658 emission).  Conversely, the new data present with a 
stronger H$\beta$ underlying absorption and He~{\sc i}~$\lambda$4921 
emission.  Note also what may be N~{\sc v} recombination emission at 
4604,20~\AA\ and at 4945,50~\AA\ (rest 
wavelengths of all four are marked).}
\label{fig:heii}
\end{figure*}

   In terms of the degree of ionization implied by the optical recombination 
lines, the environment of this binary is not quite as extreme as the 
supersoft sources.  Empirically, detected X-rays and, now, optical O~{\sc vi} 
emission seem to be wanting.

\section{Estimation of the carbon abundance and C/O ratio}

  The strength of the C~{\sc iv} and He~{\sc ii} emission lines in the 
spectra of QU~Car are remarkable.  Responding to this, we sketch here the 
simplest quantitative model of the line emission that allows an estimate of 
the abundance of carbon relative to helium and helium relative to hydrogen.  
We assume, as discussed in the previous section, that 
C~{\sc iv}~$\lambda\lambda$5801,12 arises primarily from the C$^{4+} + e- 
\rightarrow$ C$^{3+}$ recombination cascade and that C$^{4+}$ is the most 
abundant carbon ion in the emitting region (wherever that is, precisely).  
The C$^{4+}$ ion undoubtedly co-exists with helium almost entirely ionized to 
He$^{2+}$ and hydrogen ionized to H$^{+}$.  There is empirical support for 
part of this expectation in that a comparison of the red-wing velocity widths 
of the He~{\sc ii}~$\lambda$4686 and C~{\sc iv}~$\lambda\lambda$5801,12 lines 
reveals them to be well-matched.  If the He~{\sc ii} and H~{\sc i} 
emission arises from optically-thin recombination, then we can use the 
observed C~{\sc iv}/He~{\sc ii} and He~{\sc ii}/H~{\sc i} line rations to 
estimate the relative element abundances.  Briefly, we also consider 
evidence regarding the oxygen abundance.

  First, it is necessary to consider the relative strengths of accessible
He~{\sc ii} and H~{\sc i} line emission in order to determine if it is 
optically thin.  To achieve this the optical/UV SED has to be treated as 
known, so that equivalent widths can be turned into line fluxes. In the 
absence of multicolour optical photometry or indeed flux calibration of our 
optical spectrum, we adopt the same simple power law SED, normalised at 
3000~\AA\, that was derived from the fit to IUE data in section 3.  Measured 
He~{\sc ii} and H~{\sc i} equivalent 
widths and line fluxes derived with reference to this power law fit are given 
in Table 4.  The errors assigned to the derived line flux ratios in the table 
take into account equivalent width error due to continuum placement and
the effect of an SED slope error of $\pm$0.15 in the index.  All the Balmer 
emission line fluxes are superposed on broader underlying absorption profiles 
(although this is not directly apparent at H$\alpha$): the consequence of 
this for the emission line flux ratios will be a systematic error of unknown 
bias, over and above the quoted random error. The He~{\sc ii} emission, by 
contrast, is relatively clean.  In Table 4 we give theoretical Recombination 
Case B ratios for comparison (calculated using code due to P. Storey \& 
D. Hummer).  As it is likely that the line emission has an origin in a hot 
dense accretion disk atmosphere (see discussion in Hartley et al 2002), the 
ratio ranges quoted are for $10^{11} \leq n_e \leq 10^{12}$ cm$^{-3}$ and
for $T_e = 30000$~K (H~{\sc i}) or $30000 \leq T_e \leq 100000$~K 
(He~{\sc ii}).  

\begin{table*}
\caption{Emission line equivalent widths, fluxes and relevant recombination
data.  The continuum fluxes are calculated using the $\gamma = -2.25$ power
law SED derived in section 3.  The line fluxes are simply the product of
these with the equivalent widths in the previous column.    The effective
recombination coefficients given in the last column are taken from Table 14
in Kingsburgh et al (1995).}
\label{tab_emlines}
\begin{tabular}{lrccrrc}
\hline
Transition & Equivalent & Continuum  & Line & \multicolumn{2}{c}{Line Flux Ratio} & Effective recombination\\
           & Width ($-$\AA ) & Flux  & Flux & derived from & Case B & coefficient \\
 & & ergs s$^{-1}$ cm$^{-2}$ \AA$^{-1}$ & ergs s$^{-1}$ cm$^{-2}$ & observation & recombination & cm$^{3}$~s$^{-1}$ \\     
\hline
H$\alpha$ & 4.5$\pm$0.2 & 8.8$\times$10$^{-14}$ & 4.0$\times$10$^{-13}$ & 
  2.1$\pm$0.3 & 2.7$\rightarrow$2.8 & \\
H$\beta$ & 1.1$\pm$0.1 & 1.7$\times$10$^{-13}$ & 1.9$\times$10$^{-13}$ &
  1 & 1 & \\
H$\gamma$ & 0.5$\pm$0.1 & 2.2$\times$10$^{-13}$ & 1.1$\times$10$^{-13}$ &
  0.58$\pm$0.14 & 0.5$\rightarrow$0.52 & \\
 & & & & & \\
He~{\sc ii} $\lambda$1640 & 3.0$\pm$0.5 & 2.1$\times$10$^{-12}$ & 
 6.3$\times$10$^{-12}$ & 12.4$\pm$3.2 & 9.4$\rightarrow$10.6 & 2.36$\times$10$^{-24}$\\
He~{\sc ii} $\lambda$4686 & 2.7$\pm$0.2 & 1.9$\times$10$^{-13}$ & 
 5.1$\times$10$^{-13}$ & 1 & 1 & 2.38$\times$10$^{-25}$ \\
He~{\sc ii} $\lambda$5411 & 0.31$\pm$0.03 & 1.4$\times$10$^{-13}$ &
 4.3$\times$10$^{-14}$ & 0.084$\pm$0.011 & 0.087$\rightarrow$0.12 & 
 2.20$\times$10$^{-26}$ \\
 & & & & & & \\
C~{\sc iv} $\lambda\lambda$5801,12 & 1.3$\pm$0.2 & 1.2$\times$10$^{-13}$ &
 1.5$\times$10$^{-13}$ & & & 1.39$\times$10$^{-24}$ \\
 & & & & & & \\
O~{\sc iv} $\lambda$1342 & $\lesssim$0.1 & 3.3$\times$10$^{-12}$ 
 $\lesssim$3.3$\times$10$^{-13}$ & & & & 2.63$\times$10$^{-23}$ \\
\hline    
\end{tabular}
\end{table*}

  The impression to be gained from the comparison in Table 4 is that the 
Balmer decrement is a little flatter than Case B expectations, whilst the 
He~{\sc ii} ratios fit in quite well with them.  These data indicate that we 
can expect a fair estimate of $n(C^{4+})/n(He^{2+})$ from a recombination 
analysis of the C~{\sc iv}~$\lambda\lambda$5801,5812 and 
He~{\sc ii}~$\lambda$5411 lines, but the same approach using e.g. He~{\sc ii} 
$\lambda$4686 and H$\beta$ may over-estimate $n(He^{2+})/n(H^+)$.

   If the C~{\sc iv}~$\lambda\lambda$5801,5812 and He~{\sc ii}~$\lambda$5411 
emission lines form by recombination of respectively C$^{4+}$ and He$^{2+}$, 
the ratio of their fluxes is related to the recombining ion number densities 
as follows:
\begin{equation}
\frac{n(C^{4+}) \alpha_{eff, 5801,12}}{n(He^{2+}) 
        \alpha_{eff, 5411}} = \frac{F_{5801,12}}{F_{5411}}
        \frac{5806}{5411}
\end{equation}
The origins of this expression and also the effective recombination
co-efficients $\alpha_{eff, \lambda}$ we use here may be found in the
work of Kingsburgh et al (1995, Table 14 and preceding text).   The 
co-efficients they give are for an electron density, $n_e = 10^{11}$ 
cm$^{-3}$ and temperature $T_e = 50000$~K.  If, as seems plausible, the 
observed line emission in QU~Car's spectrum arises in a disk chromosphere, 
these happen to be quite appropriate conditions.  For the 5411\AA\ and 
5801,12~\AA\ lines the respective recombination co-efficients are 
$2.20\times 10^{-26}$ and $1.39\times 10^{-24}$ cm$^{3}$~s$^{-1}$.  Using 
these and further relevant data from Table 4, we obtain from equation (1):
\begin{equation}
       \frac{n(C^{4+})}{n(He^{2+})} = 0.06 \pm 0.015
\end{equation}

The error in the above expression is derived solely from the equivalent width 
uncertainties (Table 4).  Because the recombination lines compared are
close together in wavelength, the impact of the adopted SED on the derived 
relative abundance is negligible.  Potentially greater sources of error
arise from the adopted ($n_e$, $T_e$) that goes with the use of the Kingsburgh
et al (1995) effective recombination co-efficients, and from the assumed
recombination model for the 5801,5812\AA\ emission.  Neither of these is
easy to quantify.  We suspect that the choices of density and temperature will
not amount to more than a factor of $\sim$2 uncertainty -- after all to
a recombined electron in high principal quantum number states, the C$^{3+}$
and He$^{+}$ ions do not seem very different (the fact lying behind the 
near-coincidences in wavelength between so many highly-excited C{\sc iv} and
He{\sc ii} lines).  

The bigger systematic error may be more to do with the implied model for the 
level populations involved in the C{\sc iv}~$\lambda\lambda$5801,12 doublet 
transition itself.  The upper level of the transition, 1s$^2$3p ($^2$P$^o$) 
can be directly excited from the ground state via the resonant transition at 
312\AA\ as well as populated by recombination.  This fact also means that 
there may be non-negligible opacity in the 5801,12~\AA\ doublet when
the 312~\AA\ resonance is very opaque -- possibly explaining the hint of 
blueshifted absorption in the observed profile.  These considerations show
that it is not obvious whether the emission in this C~{\sc iv} transition is 
raised above or below the pure recombination value.  Indeed the experience of 
Kingsburgh et al (1995) with this line for WO-star abundance determinations 
is that such complications can cause both over- and under-estimation by a 
factor of a few.  

For the present, we can check for consistency between the emission in
5801,12~\AA\ line and that in other C~{\sc iv} lines in the observed
wavelength range.  C~{\sc iv}~$\lambda$4658 is undoubtedly present in QU~Car's
spectrum, but of course it is blended into the $\sim$4650~\AA\ emission 
feature along with C~{\sc iii} and some N~{\sc iii}.  The effective 
recombination co-efficent for this line is 8.32$\times$10$^{-25}$ 
cm$^{3}$~s$^{-1}$ -- more than half that for C~{\sc iv}~$\lambda\lambda$5801,12
(Table 4).  From a multiple Gaussian fit to the 4650~\AA\ blend, that imposes 
the correct relative wavelengths of the contributing N{\sc iii}, C{\sc iii} 
and C{\sc iv} multiplet components and the relative emission
strengths (within each multiplet) expected for optically-thin gas, we find 
that the C{\sc iv}~$\lambda$4658 multiplet could contribute as much as a third 
of the observed flux in the blend (i.e. an equivalent width of up to 
$\sim$0.9~\AA ).  Based on the observed equivalent width of the 5801,12~\AA\ 
doublet, we would have predicted half this much!  So one might take from this
that the $n(C^{4+})/n(He^{2+})$ ratio is underestimated using the
5801,12~\AA\ doublet.  But to counterbalance this, there is not a detection
of the weak C{\sc iv}~$\lambda$5470 line when perhaps there should have been, 
just -- this could imply an error of a factor of 2 or more in the opposite 
sense.

Accordingly we recognise that the derived $n(C^{4+})/n(He^{2+})$ may be 
wrong by a factor of 2 to 4.  At the same time, it is possible that
the He~{\sc ii} line emission forms in a larger volume than the C~{\sc iv}
line emission, with the implication the ratio given as expression (2) could be 
a lower bound on the C/He abundance ratio.  For the moment, our conservative
estimate is that C/He is {\em at least} $\sim$0.015.  As a comparison, we 
note that the solar value for this ratio is 0.004 (Sofia \& Meyer 2001, for 
10\% He by number relative to H).  The reason why carbon line emission from a 
range of ion stages is so plentiful in the UV/optical spectrum of QU~Car is 
now evident (see e.g. Figure 9 in Hartley et al 2002, and GP82).

    Over a wide range of temperatures and densities the ratio between
Case B effective recombination co-efficients for the He~{\sc ii}~$\lambda$4686 
and H$\beta$ lines is within 10 percent of 9.  Note that this means that a 
normal helium abundance of 1/10 by number relative to hydrogen converts to 
rough equality in the equivalent widths of He~{\sc ii}~$\lambda$4686 and 
H$\beta$ emission (if He$^{2+}$ and H$^{+}$ are the dominant ion stages).  
Table 4 (and Figure 5) shows that in our optical spectrum the 
He~{\sc ii}~$\lambda$4686 equivalent width is about 2.5 times higher than 
that of H$\beta$ -- implying a helium over-abundance of a similar order.  
Indeed, for clear nucleosynthetic reasons, it could be a problem to find 
evidence of carbon enrichment without some helium enrichment also.  If 
optical depth is implicated in the slight flattening of the Balmer 
decrement relative to Recombination Case B, or if Case A is the better
approximation to apply, this factor of $\sim$2.5 would have to be
lowered.  Putting this together with the estimated C/He ratio, [C/H] in 
QU~Car could be anything between 4 and 40.  

    Finally, as it will be helpful to form an impression of the C/O ratio
we look for a rough measure of the oxygen abundance (relative to helium).
The non-detections of O{\sc vi}~$\lambda$5290 and O{\sc v}~$\lambda$5590
can be re-expressed (conservatively) as equivalent width upper limits of 
$\sim$0.03~\AA .  Through the versions of equation 1 that relate the
number densities of O$^{6+}$ and O$^{5+}$ to the He$^{2+}$ number density,
using the 5290\AA , 5590\AA\ and 5411\AA\ lines, we can derive the following
upper limiting ion ratios:
\begin{eqnarray}
    n(O^{6+})/n(He^{2+}) & < & 0.002 \\
    n(O^{5+})/n(He^{2+}) & < & 0.0005
\end{eqnarray}
To complete the picture, it is necessary to find an estimate of the O$^{4+}$
abundance since this may very well be the dominant oxygen ion in the volume
where C$^{4+}$ and He$^{2+}$ are found.  In the merged HST spectrum 
(Figure 1) there is the slightest hint of O{\sc iv}~$\lambda$1342 in emission.
Estimates of its equivalent width and flux are quoted in Table 4.
Measurement of this emission is bound to be an uncertain business both 
because it is extremely weak and also because the emission may very well be
close to cancelling underlying absorption (i.e. assuming line formation 
solely by recombination is likely to be an over-simplification).  But it
is the best we can do for now.  Using the data in Table 4 for the
O~{\sc iv}$\lambda$1342 and He~{\sc ii}~$\lambda$1640 lines, we obtain 
\begin{equation}
\frac{n(O^{4+})}{n(He^{2+})} \lesssim 0.004 
\end{equation}
So, if it is assumed that the detection (just) of the 1342\AA\ line in 
emission gives a useful constraint on the O/He number ratio, it follows that 
the C/O abundance ratio is in the region of 10.  Whilst this is a very
uncertain figure, it amounts to a strong indication that this ratio exceeds
unity.

\section{Summary and discussion}
    
   In this study, two significant findings have emerged.

   First, UV interstellar absorption line data have revealed evidence of a 
blueshifted diffuse cloud components at an LSR velocity of $-14$ km s$^{-1}$ 
(to within 2 km~s$^{-1}$).  Interpreted in terms of a standard galactic 
rotation model, this tells us that the minimum distance to QU~Car is about 
1.8~kpc, rather than $\sim$0.5~kpc as estimated by GP82.  Nevertheless the 
fact that the H{\sc i} interstellar column towards QU Car is about a third 
the total for the Galaxy along the same siteline, reassures that it is located
in the Galaxy.  In fact, if the distance to QU~Car actually is $\sim 2$~kpc,
it would be at much the same distance from the Sun as the Carina spiral arm 
at the same galactic longitude (Grabelsky et al 1987).  

   Second, a newly-obtained intermediate dispersion optical spectrum of 
QU~Car has been found to contain no O~{\sc vi} line emission and, yet, 
unusually strong C~{\sc iv}~$\lambda\lambda$5801,12 emission is present.  It 
has long been known that the $\sim$4650~\AA\ emission feature is also very 
strong and appears to be dominated by carbon, rather than nitrogen, line 
emission (GP82).  Starting from the strength of the 
C~{\sc iv}~$\lambda\lambda$5801,12 emission compared to 
He~{\sc ii}~$\lambda$5411 emission, we have shown that there is
very likely significant enrichment of carbon in QU~Car's emission 
line region (with [C/H] in the range 4 to 40).  Furthermore, the near
absence of any oxygen line emission from the UV and optical spectrum
suggests the C/O abundance ratio exceeds unity.

   We now consider the implications for the status of QU~Car if its distance 
from the Sun is $\sim$2~kpc or more.  It was shown in section 3 that this
demands a high luminosity, $L_{{\rm bol}} \sim 10^{37}$ ergs s$^{-1}$,
and hence a high mass accretion rate, $\mdot_a \sim 10^{-7}$ M$_{\odot}$ 
yr$^{-1}$ (assuming a white dwarf accretor).  
 
   Another way of visualising this is to scale QU~Car with 
respect to the most famous and brightest of dwarf novae, SS~Cyg.  This is a 
useful comparison also because SS~Cyg is a low-inclination binary, just
as we believe QU~Car to be (see Hartley et al 2002).  Parallax observations 
have recently led to an upward revision of the distance to SS~Cyg placing it 
at a distance of 166~pc (Harrison et al 1999).  Schreiber \& G\"ansicke (2002) 
have re-evaluated the outburst mass accretion rate, using the revised 
distance: at $V = 8.5$, they associate with it the distinctly high mass 
transfer rate of $(5.7 \pm 2.4) \times 10^{-8}$ M$_{\odot}$~yr$^{-1}$ (for 
the widely accepted white dwarf mass of 1.19~M$_{\odot}$).  After correcting 
for reddening, the apparent magnitudes of QU~Car and SS~Cyg imply that the 
former is 12 times fainter than the latter.  This means QU~Car is 
intrinsically 2.6 $\times$ D$^2$ brighter in the $V$ band, where $D$ is the 
distance to QU~Car expressed in kpc.  Hence at $\sim$2~kpc, QU Car is  
10 times more luminous than SS~Cyg in the $V$ band, and possibly powered by a 
mass transfer rate as high as $\sim 6 \times 10^{-7}$ M$_{\odot}$~yr$^{-1}$ 
(scaling from Schreiber \& G\"ansicke's calculation).  

   The clear empirical restraint on relabeling QU~Car as a supersoft source is 
that there is only an upper limit on the soft X-ray flux (Verbunt et al 1997).
A low X-ray luminosity is also consistent with the absence of O~{\sc vi}
recombination lines in the spectra we have obtained.  Nevertheless, given 
that `X-ray off' states have been recorded for supersoft sources 
(e.g. RX~J0513.9$-$6951, Reinsch et al 1996; CAL~83, Greiner \& DiStefano 
2002), this difference might 
diminish.  A wavelength domain in which QU~Car bears some resemblance to 
known supersofts is in the ultraviolet.  G\"ansicke et al (1998) have described
the UV spectra of RX~J0513.9$-$6951 and CAL 83, and point out the presence
of N~{\sc v}~$\lambda$1240, O~{\sc v}~$\lambda$1371 and 
He~{\sc ii}~$\lambda$1640 emission in both objects.  These are also prominent 
in QU~Car, with the difference that the N~{\sc v} and O~{\sc v} features are 
in absorption -- this might also signal less extreme ionization than in
the supersoft sources.  It is unlikely that orbital inclination is 
implicated here given that all three objects are believed to be at low 
inclination (Hartley et al 2002; Cowley et al 1998).  Finally we note that
for a distance of 2~kpc the absolute magnitude, $M_V = -0.4$, of QU~Car is 
fainter than that of these same two supersofts by 1.6 and 0.9 mags.

   It remains possible that the interstellar line data mislead
in indicating the longer siteline to QU~Car, leaving the earlier-proposed 
minimum distance of 0.5~kpc intact.  At this nearer limiting distance, QU~Car
is still the most luminous nova-like variable known.  Indeed using the
scaling to SS~Cyg again, at $D = 0.5$~kpc, the mass transfer rate 
would be expected to be in the region of $4\times 10^{-8}$ 
M$_{\odot}$~yr$^{-1}$ (see also Table 2, section 3).

   Hitherto, on the rare occasions that deviations from cosmic abundances 
in CV have been measured and documented, they have been in the 
sense of enhancements due to CNO-cycle burning (i.e. mainly nitrogen 
enhancement, resulting from conversion of carbon and perhaps oxygen: see 
Long 2000, Marsh et al 1995).  QU~Car appears 
to be the first recognised instance of enhancement due to helium burning.  
This would not be remarkable if the line emission concerned could be 
associated with the accreting white dwarf.  We are reluctant to propose this 
since the character of QU~Car's UV emission line spectrum points toward an 
accretion disk origin (Hartley et al 2002).  

   The carbon 
enhancement has then to be located in the envelope of the companion star.  
Could the companion star be a CO white dwarf?  There can be two objections 
to this: (i) the published orbital period of 10.9 hours (GP82) is too long, 
(ii) a double-degenerate CV would not be expected to be so bright (even
for the shorter minimum distance of 500~pc).  To counter the first objection 
it can be proposed that the time series sampling of GP82 was too coarse to 
pick up the short ($\sim$~1000 sec) periods expected for double-degenerate 
systems.  Certainly, a re-examination of QU~Car's binary parameters is 
warranted, given the extraordinary character of this binary and the ragged
character of the phase-folded radial velocity curve (see GP82).  The second 
objection currently has a simple empirical basis in that the absolute 
magnitude of the brightest of the known AM CVn systems (AM~CVn itself) is
thought to be $M_V \sim 9$ (Warner 1995; see also Nasser, Solheim \& 
Semionoff 1991).  A third problem for this scenario is the undeniable 
presence of hydrogen line emission and absorption in QU~Car's optical 
spectrum.

   Accepting, for the moment, that the orbital period really is 10.9~hours
as determined by GP82, the implication from the period-mean density 
relation (e.g. Eggleton 1983) is that the companion star should have the mean 
density of a mid-F main sequence star (or of a star evolving away from a 
later-type position on the main sequence).  Near main sequence stars
cannot exhibit the pronounced carbon enhancement that has been discovered 
here.  At higher luminosity, the abundance patterns that have been determined 
for barium and CH giants will not fit either, given that their carbon 
enrichment is against a backdrop of overall reduced metallicity, with the
result that C/H or C/He is typically still less than in the Sun (see 
Barbuy et al 1992, Vanture 1992).  

   Nevertheless there is one class of not-so-bright giant that does come 
close to matching the abundance pattern deduced for QU~Car -- the early-type 
R stars.  These are carbon stars now known to have absolute magnitudes that 
place them on the HRD as red clump giants ($M_K \simeq -2$, with $V-K$ in the 
range from 2 to 4, Knapp, Pourbaix \& Jorissen 2001).  In the R-star sample 
of Dominy (1984), carbon enhancements up to $\sim$6 times solar are found 
with little change or a small drop in oxygen abundance.  To accommodate such 
a companion within QU~Car, the binary needs to be at or beyond the larger 
minimum distance discussed here.  Specifically, at 2~kpc the absolute visual 
magnitude of QU~Car is -0.4, to be compared with the same quantity for 5 
R stars from Wallerstein and Knapp (1998) ranging from -0.25 up to 4.17.  
It would seem that the fainter half of this range is consonant with the 
non-detection of the companion star in QU~Car's optical spectrum.   If 
further work can substantiate this concept, it could lend support to the 
growing opinion that the abundance patterns in early-type R stars have to do 
with outward mixing after the core helium flash (see discussion in Knapp et al 
2001).  The alternative idea that the failure to detect binary motion in these 
carbon-rich stars points to their being coalesced binaries (see e.g. 
McClure 1997) sits awkwardly with the fact that QU~Car remains apparent as a 
binary in which the mass transfer is {\em away} from the putative R star.  
However work will have to be done to see whether an early-type R star in a 
high mass transfer rate system is a plausible consequence of binary star 
evolution.  In this context, the 10.9~hr orbital period may prove too
short for comfort -- unless the R star can have already shed much of its 
envelope.

    If the distance to QU~Car is closer to $\sim$500~pc, we are unable to
suggest a plausible identity for the companion star.  A dwarf carbon-star
companion probably can be ruled out since the overall abundance pattern in
such objects is that of a reduced-metallicity Population II star, like the 
CH and Ba giants (see Wallerstein \& Knapp 1998).  It is also doubtful that 
pollution of the companion star by carbon-rich debris from unrecorded nova 
explosions in the past is a viable explanation -- if the white dwarf has 
accreted from an unexceptional companion star, nucleosynthesis in the nova 
outburst will most likely yield nitrogen-enhanced ejecta with C/O $<$ 1
(see Gehrz et al 1998).   Putting all the arguments together, we conclude that
the apparent abundance peculiarity in QU~Car has to be seen as favouring the 
picture in which this binary is distant and luminous.  

   We end with the remark that in carrying out this piece of work we have
come to appreciate the paucity of published CV optical spectra (particularly 
outside the blue range).  The only comprehensive optical spectroscopic survey 
of CVs that we have been able to draw on is the work of Williams (1983).  
Among the spectra presented there it can be seen that only T~Pyx and V~Sge 
bear a family resemblance to QU~Car, particularly in terms of the strength of
He~{\sc ii} lines relative to those of H~{\sc i}.  Furthermore in the case of 
V~Sge one may, perhaps uniquely, just make out from the plot of its spectrum
the C~{\sc iv}~$\lambda\lambda$5801,5812 blend in emission -- certainly 
Williams did not find this feature often enough to consider quoting 
measurements of it in tables.  Referring to the study of Herbig et al (1965) 
on this well-known and perplexing binary, we find a report of data on the 
C~{\sc iv} transition giving an equivalent width 1.3 times that of the 
nearby He{\sc ii}~$\lambda$5411 line.  In QU~Car this ratio is close to 4! 
If V~Sge is bizarre, QU~Car is also.  Objects such as these are testaments
to the propensity that nature has to try out all the options -- and in binary
star evolution there appears to be especially many. 

\section*{Acknowledgments}

JED would particularly like to thank Mike Barlow for drawing attention to the 
carbon and oxygen recombination coefficients in the work of Kingsburgh et al 
(1995) and for discussion in connection with their use.  Thanks
are also due to Dave Kilkenny for putting JED and JvW in touch and helping
set up the SAAO observations.  LEH acknowledges the award of a studentship 
from the Particle Physics and Astronomy Research Council (PPARC) of the 
United Kingdom.  Support for HST proposal number G0-8279 was provided by NASA 
through a grant from the Space Telescope Science Institute, which is operated 
by the Association of Universities for Research in Astronomy,
Incorporated, under NASA contract NAS5-26555.  In addition, this paper is 
based on data obtained from the South African Astronomical Observatory.  Use 
was also made of the atomic line list (http://www.pa.uky.edu/~peter/atomic/)
due to Peter van Hoof.  We thank the referee, Frank Verbunt, for his useful
comments.

\end{document}